\documentclass{ws-ijmpa}
\begin{document}

\markboth{Alejandro Gangui}
{Early Universe Sources for Cosmic Microwave Background Non-Gaussianity}

\catchline{}{}{}

\title{Early Universe Sources for 
CMB
Non-Gaussianity}
\author{\footnotesize ALEJANDRO GANGUI}
\address{Instituto de Astronom\'{\i}a y F\'{\i}sica del Espacio,
         1428 Buenos Aires, Argentina, and   \\
         Departamento de F\'{\i}sica, Universidad de Buenos Aires, 
         1428 Buenos Aires, Argentina.}

\maketitle

\pub{Received (Day Month Year)}{Revised (Day Month Year)}

\begin{abstract}

In the framework of inflationary models with non-vacuum initial states
for cosmological perturbations, we study non-Gaussian signatures on
the cosmic microwave background (CMB) radiation produced by a
broken-scale-invariant model which incorporates a feature at a
privileged scale in the primordial power spectrum.

\end{abstract}

\section{Introduction}

The common belief that the CMB is Gaussian distributed can be directly
traced back to the generic assumption that the quantum fluctuations of
the inflaton field are placed in the vacuum state
~\cite{Grishchuk-Martin}. Relaxing this assumption might lead to
detectable signatures in various astrophysical tests, most
interestingly in future CMB and large-scale structure observations.
In this note, we study CMB non-Gaussian signatures predicted within
inflationary models with non-vacuum initial states for cosmological
perturbations. The model incorporates a privileged scale, which
implies the existence of a feature in the primordial power
spectrum. The model predicts a vanishing three-point correlation
function for the CMB temperature anisotropies~\cite{MRS}. We here
focus on the first non-vanishing moment, the CMB four-point function
at zero lag, namely the kurtosis, and compute its expected value for
different locations of the primordial feature in the spectrum, as
suggested in the literature to conform to observations of large scale
structure~\cite{silk,einasto}.

\section{Two-point correlation function for non-vacuum initial states}

We consider non-vacuum states for the cosmological perturbations of
quantum mechanical origin. Let ${\cal D}(\sigma )$ be a domain in
momentum space, such that if ${\bf k}$ is between $0$ and $\sigma $,
the domain ${\cal D}(\sigma )$ is filled by $n$ quanta, while
otherwise ${\cal D}$ contains nothing.  The state $\vert \Psi
_1(\sigma ,n)\rangle $ is defined by
\begin{equation}
\label{defpsi1}
|\Psi _1(\sigma ,n)\rangle \equiv
\prod _{{\bf k} \in {\cal D}(\sigma )}
\frac{(c_{\bf k}^{\dagger })^n}{\sqrt{n!}} |0_{\bf k}\rangle
\bigotimes _{{\bf p}\not\in {\cal D}(\sigma )}|0 _{\bf p}
\rangle
=\bigotimes _{{\bf k} \in {\cal D}(\sigma )}|n_{{\bf k}}\rangle
\bigotimes _{{\bf p} \not\in {\cal D}(\sigma )}|0_{\bf p}\rangle .
\end{equation}
The state $|n_{{\bf k}}\rangle$ is an $n$-particle state satisfying,
at conformal time $\eta=\eta_{\rm i}$: $c_{\bf k}|n_{{\bf
k}}\rangle=\sqrt{n}|(n-1)_{{\bf k}}\rangle$ and $c^{\dag }_{\bf
k}|n_{{\bf k}}\rangle=\sqrt{n+1}|(n+1)_{{\bf k}}\rangle$.  We have the
following property
\begin{equation}
\label{proppsi1}
\langle \Psi _1(\sigma ,n)|\Psi _1(\sigma' ,n')\rangle=
\delta (\sigma -\sigma ')\delta _{nn'}.
\end{equation}           
It is clear from the definition of the state $|\Psi _1\rangle $ that
the transition between the empty and the filled modes is sharp.  In
order to ``smooth out'' the state $|\Psi _1\rangle $, we consider a
state $|\Psi _2\rangle $ as a quantum superposition of $|\Psi
_1\rangle $.  In doing so, we introduce an, a priori, arbitrary
function $g(\sigma ;k_{\rm b})$ of $\sigma$. The definition of the
state
$\vert \Psi_2(n,k_{\rm b})\rangle$
\begin{equation}
\label{defpsi2}
|\Psi _2(n,k_{\rm b})\rangle \equiv \int _0^{+\infty }{\rm d}\sigma
g(\sigma ;k_{\rm b}) \vert \Psi _1(\sigma ,n)\rangle ,
\end{equation}
where $g(\sigma ;k_{\rm b})$ is a given function which defines the
privileged scale $k_{\rm b}$. We assume that the state is normalized
and therefore $\int _0^{+\infty }g^2(\sigma ;k_{\rm b}){\rm d}\sigma
=1$. In the
state $\vert \Psi _1(\sigma ,n)\rangle $, for any domain ${\cal D}$
one has   
\begin{eqnarray}
\langle \Psi _1(\sigma ,n)|c_{\bf p}c_{\bf q}|
\Psi _1(\sigma ,n)\rangle
&=& \langle \Psi _1(\sigma ,n)|c_{\bf p}^{\dag}
c_{\bf q}^{\dag}|\Psi _1(\sigma ,n)\rangle =0, \\
\langle \Psi _1(\sigma ,n)|c_{\bf p}
c_{\bf q}^{\dag}|\Psi _1(\sigma ,n)
\rangle
&=& n{\rm \delta }({\bf q}\in {\cal D}){\rm \delta }({\bf p}-{\bf q})
+{\rm \delta }({\bf p}-{\bf q}), \\
\langle \Psi _1(\sigma ,n)|c_{\bf p}^{\dag }
c_{\bf q}|\Psi _1(\sigma ,n)
\rangle &=& n{\rm \delta }({\bf q}\in {\cal D}){\rm \delta }({\bf
p}-{\bf q}).
\end{eqnarray}
In these formulas, ${\rm\delta }({\bf q}\in {\cal D})$ is a function
that is equal to 1 if ${\bf q}\in {\cal D}$ and 0 otherwise. These
relations will be employed in the sequel for the computation of the
CMB temperature anisotropies for the different non-vacuum initial
states.

\subsection{Two-point function of the CMB temperature anisotropy}

The spherical harmonic expansion of the cosmic microwave background
temperature anisotropy, as a function of angular position, is given by
\begin{equation}
\label{dTT}
\frac{\delta T}{T}({\bf e})=\sum _{\ell m}a_{\ell m}
Y_{\ell m}({\bf e})
\qquad
{\rm with}
\qquad
a_{\ell m}=\int
{\rm d}\Omega _{{\bf e}}\frac{\delta T}{T}({\bf e})Y_{\ell m}^*({\bf
e}).
\end{equation}
As we are
interested in a non-Gaussian signature of primordial origin we will be
focusing on large angular scales, for which the main contribution
to the temperature anisotropy is given by the Sachs-Wolfe effect,
namely, 
${\delta T}/{T}({\bf e})\simeq
({1}/{3})\Phi [\eta _{\rm lss},{\bf e}(\eta _0-\eta _{\rm lss})],
$
where $\Phi (\eta ,{\bf x})$ is the Bardeen potential, while $\eta _0$
and $\eta _{\rm lss}$ denote respectively the conformal times now and
at the last scattering surface. Note that the previous expression is
only valid for the standard Cold Dark Matter model (sCDM). In general,
we might also be interested in the case where a cosmological
constant is present ($\Lambda $CDM model) since this seems to be
favored by recent observations. Then, the integrated Sachs-Wolfe
effect plays a non-negligible role on large scales and the expression
giving the temperature fluctuations is not as simple as the previous
one.                           

In the theory of cosmological perturbations of quantum mechanical
origin, the Bardeen variable becomes an operator, and its expression
can be written as
\begin{equation}
\label{Bop}
\Phi (\eta ,{\bf x})=\frac{\ell_{\rm Pl}}{\ell_0}\frac{3}{4\pi }
\int {\rm d}{\bf k}
\biggl[c_{\bf k}(\eta _{\rm i})f_k(\eta )e^{i{\bf k}\cdot {\bf x}}
+c_{\bf k}^{\dag }(\eta _{\rm i})f_k^*(\eta )e^{-i{\bf k}
\cdot {\bf x}}\biggr]~,
\end{equation}
where $\ell_{\rm Pl}=(G\hbar)^{1/2}$ is the Planck length. In the
following, we will consider the class of models of power-law inflation
since the power spectrum of the fluctuations is then explicitly
known. In this case, the scale factor reads $a(\eta )=\ell _0\vert
\eta \vert ^{1+\beta }$, where $\beta \le -2 $ is a priori a
free parameter. However, in order to obtain an almost scale-invariant
spectrum, $\beta $ should be close to $-2$. In the previous expression
of the scale factor, the quantity $\ell _0$ has the dimension of a
length and is equal to the Hubble radius during inflation if $\beta
=-2$.  The parameter $\ell _0$ also appears in Eq.~(\ref{Bop}). 
The mode function $f_k(\eta
)$ of the Bardeen operator is related to the mode function $\mu
_k(\eta )$ of the perturbed inflaton through the perturbed Einstein
equations. In the case of power-law inflation and in the long
wavelength limit, the function $f_k(\eta )$ is given in terms of the
amplitude $A_{_{\rm S}}$ and the spectral index $n_{\rm s}$ of the
induced density perturbations by    
\vspace*{-13pt}
\begin{equation}
\label{sind}
k^3|f_k|^2=A_{_{\rm S}}k^{n_{\rm s}-1}~.
\end{equation}
Using the Rayleigh equation and the completeness relation for the
spherical harmonics 
and after some algebra we get 
\begin{equation}
\label{linkaphi}
a_{\ell m}=\frac{\ell_{\rm Pl}}{\ell_0}
e^{i\pi \ell /2}\int {\rm d}{\bf k}
\biggl[c_{\bf k}(\eta _{\rm i})f_k(\eta )
+c_{-{\bf k}}^{\dag }(\eta _{\rm i})f_k^*(\eta )\biggr]
j_{\ell }[k(\eta _0-\eta _{\rm lss})]Y_{\ell m}^*({\bf k})~.
\label{alm}
\end{equation}
At this point we need to somehow restrict the shape of the domain
${\cal D}$. We assume that the domain only restricts the modulus of
the vectors, while it does not act on their direction. Then, from
Eq.~(\ref{alm}), one deduces                            
\begin{eqnarray}
\label{clpsi1}
\langle \Psi _1(\sigma ,n)\vert a_{\ell _1m_1}a_{\ell _2m_2}^*
\vert \Psi _1(\sigma ,n)\rangle
&=& \frac{\ell_{\rm Pl}^2}{\ell_0^2}\biggl[C_{\ell _1}
+2nD_{\ell _1}^{(1)}(\sigma )\biggr]
\delta _{\ell _1\ell_2}\delta _{m_1m_2}~,
\end{eqnarray}
\vspace*{-13pt}
\begin{equation}
\label{defdl1}
D_{\ell }^{(1)}(\sigma )\equiv \int _0^{\sigma }j_{\ell }^2[k(\eta
_0-\eta _{\rm lss})] k^3|f_k|^2\frac{{\rm d}k}{k} 
\end{equation}
Thus, the multipole moments
$C_{\ell }^{(1)}$, in the state $|\Psi _1\rangle $, are given by
$
C_{\ell }^{(1)}(\sigma )=C_{\ell }+2nD_{\ell }^{(1)}(\sigma )~,
$
where $C_{\ell }$ is the ``standard'' angular power spectrum, i.e., the multipole
obtained in the case where the quantum state is the vacuum, i.e.,
$n=0$. Let us calculate the same quantity in the state $|\Psi
_2\rangle $.
Performing a similar analysis as the above one, we find~\cite{GMS}  
\begin{eqnarray}
\label{clpsi2final}
\langle \Psi _2(n,k_{\rm b})\vert a_{\ell _1m_1}a_{\ell _2m_2}^*
\vert \Psi _2(n,k_{\rm b})\rangle 
&=&
\frac{\ell_{\rm Pl}^2}{\ell_0^2}\biggl[C_{\ell _1}
+2nD_{\ell _1}^{(2)}\biggr]\delta _{\ell _1\ell_2}\delta _{m_1m_2}~,
\end{eqnarray}
\vspace*{-16pt}
\begin{equation}
\label{defdl2}
D_{\ell }^{(2)}
=\frac{\pi }{2}A_{_{\rm S}}
\int _0^{+ \infty}J_{\ell +1/2}^2(k)
\bar{h}(k)k^{n_{\rm S}-3}{\rm d}k
\end{equation}
where, to reach this eqn, we defined $g^2(\sigma ;k_{\rm b})\equiv {\rm d}h/{\rm d}\sigma $ [we
will see below that this function $h(k_{\rm b})$ cannot
be arbitrary] and we integrated by parts, and then we defined 
$\bar{h}(k) \equiv h(\infty )[1-h(k)/h(\infty)]$. In this 
we have not assumed anything on $h(\infty )$ or
$h(0)$. We see that the relation $g^2(k)\equiv {\rm d}h/{\rm d}k$
requires the function $h(k)$ to be monotonically increasing with
$k$. It is interesting that, already at this stage of the
calculations, very stringent conditions are required on the function
$h(k)$ which is therefore {\it not} arbitrary. This implies that the
function $\bar{h}(k)$ which appears in the correction to the multipole
moments is always positive, vanishes at infinity and is monotonically
decreasing with $k$. An explicit profile for $\bar{h}(k)$ is given
in Fig.~\ref{fig1}. 
The total power spectrum of the Bardeen potential can be written
as                           
\begin{equation}
\label{psba}
k^3\vert \Phi _k\vert ^2\propto A_{_{\rm S}}k^{n_{\rm S}-1}
\biggl\{1+2nh(\infty )\biggl[1
-\frac{h(k)}{h(\infty )}\biggr]\biggr\}=A_{_{\rm S}}k^{n_{_{\rm S}}-1}
[1+2n\bar{h}(k)].
\end{equation}
Observations indicate that $n_{_{\rm S}}\simeq 1$ and for
simplicity we will take $n_{_{\rm S}}=1$. 
As we have seen previously, we can write the multipole moments in 
the state $|\Psi _2\rangle $ as
$C_{\ell }^{(2)}=C_{\ell }+2nD_{\ell }^{(2)}$.
Substituting the well-known expression for the $C_{\ell }$'s and the
definition of $D_{\ell }^{(2)}$ given by Eq.~(\ref{defdl2}),
one finds that the coefficients $C_{\ell }^{(2)}$ are given by 
\begin{equation}
\label{multl}
C_{\ell }^{(2)}=A_{_{\rm S}}\frac{\pi }{2}\biggl\{
{1\over 2^{3-n_{\rm s}}} {\Gamma(3-n_{\rm s})\Gamma[\ell +(n_{\rm
s}-1)/2]
\over \Gamma^2[(4-n_{\rm s})/2]\Gamma[\ell -(n_{\rm s}-5)/2]}
+2n{\bar D_{\ell }}^{(2)}\biggr\}.
\end{equation}
As a next step, one has to normalize the spectrum (need to determine
the value of $A_{\rm S}$). We choose to use the value of
$Q_{{\rm rms-PS}} = T_0 [5C_2^{(2)} / (4\pi)]^{1/2}
(\ell_{\rm Pl}/\ell_0) \sim 18 \mu K$ with $T_0=2.7K$ measured by the
COBE satellite. Thus, we compute the quadrupole and then
\begin{equation}
\label{coucouadrupole}
A_{_{\rm S}}=\frac{8}{5}
\frac{Q_{\rm rms-PS}^2}{T_0^2}\frac{\ell_0^2}{\ell_{\rm Pl}^2}
\biggl[\frac{1}{6\pi }+2n{\bar D_2}^{(2)}\biggr]^{-1},
\end{equation}
for $n_{_{\rm S}}=1$. 
The 
band power $\delta T_{\ell }$ 
gives 
\begin{equation}
\label{bandn=1}
\delta T_{\ell }=\frac{Q_{\rm rms-PS}}{T_0}
\sqrt{\frac{12}{5}}\sqrt{\frac{1+2n\pi \ell (\ell +1)
{\bar D_{\ell }}^{(2)}}{1+12n\pi
{\bar D_{2}}^{(2)}}}~.
\end{equation}
The $n$-dependence in the above expression is the correction due to
the non-vacuum initial state. We easily check that if $n=0$ the
corresponding band powers are constant at large angular scales.

Finally, we calculate the two-point correlation function at zero lag
in the state $|\Psi_2\rangle$.
Using Eqs.~(\ref{dTT}),~(\ref{clpsi2final}),
the second moment, $\mu_2$, of the distribution is given by
\begin{equation}
\label{2point}
\mu_2 \equiv
\biggl \langle \biggl[\frac{\delta T}{T}({\bf e})\biggr]^2\biggr
\rangle
=\frac{\ell_{\rm Pl}^2}{\ell_0^2}
\sum _{\ell }\frac{2\ell +1}{4\pi }C_{\ell }^{(2)} ~.
\end{equation}
Once we have reached this point, an obvious first thing to do is to
check that the two-point correlation function calculated above is
consistent with present observations.                

\subsection{Comparison with observations}      

\begin{figure}
\centerline{\psfig{file=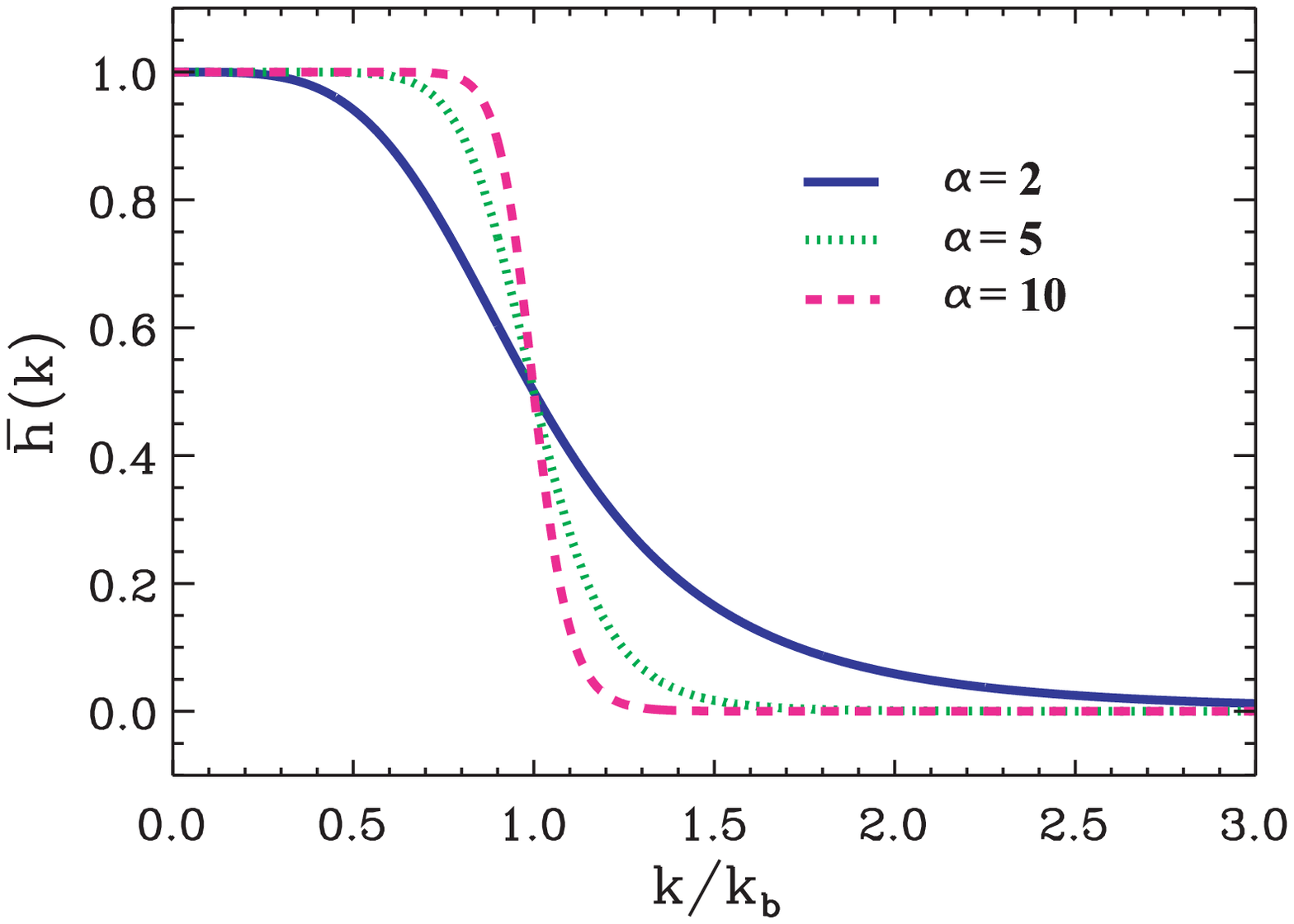,width=6cm}\psfig{file=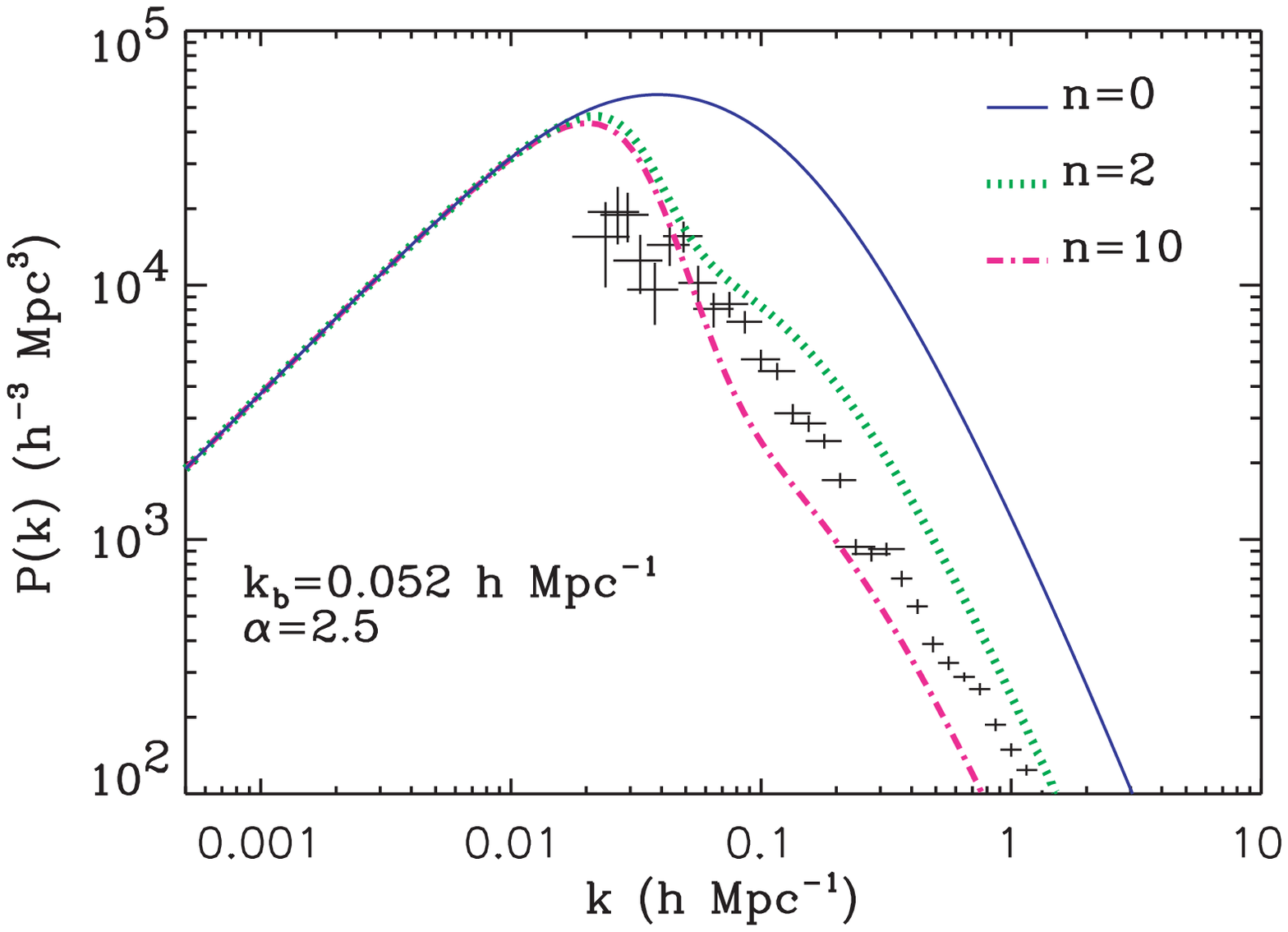,width=6.5cm}}
\vspace*{-9pt}
\caption{The function $\bar{h}(k)$ for different values of $\alpha$
(left panel).  On the right panel we show the matter power spectrum
normalized to COBE for different numbers $n$ of quanta in the initial
state. The cosmological parameters are those corresponding to the sCDM
model, namely, $h=0.65$, $\Omega _{\Lambda }=0$, $\Omega _{\rm
b}=0.05$, $\Omega _{\rm cdm}=0.95$ and $n_{_{\rm S}}=1$. The
parameters describing the non vacuum state are $k_{\rm b}^{\rm
phys}=0.052 h \mbox{Mpc}^{-1}$ and $\alpha =2.5$. The data points
represent the power spectrum measured by the PSCz survey.}\label{fig1}\end{figure}

Among the available observations that one can use to check the
predictions of theoretical models, two are key in cosmology: the CMB
anisotropy and the matter-density power spectra. 
We will not study in details all the predictions that can
be done from the two-point correlation function since our main purpose
in this note is to calculate the non-Gaussianity which is a clear
specific signature of a non vacuum state.
So, we just compute the matter power
spectrum to demonstrate that it fits reasonably well the available
astrophysical observations for some values of the free parameters. In
addition, this illustrates well the fact that, using the available
observations, we can already put some constraints on the free
parameters. 
A simple ansatz for the function $\bar{h}(k)$ is represented in
Fig.~\ref{fig1} and can be expressed as
\begin{equation}
\label{barh}
\bar{h}(k)=\frac{1}{2}\biggl[1-\tanh
\biggl(\alpha \ln \frac{k}{k_{\rm b}}\biggr)\biggr]~.
\end{equation} 
With it, the matter power spectrum today, after taking into account
the transfer function $T(k)$ which describes the evolution of the
Fourier modes inside the horizon, can be written
\begin{equation}
\label{specmod}
P(k)=T^2(k)
\frac{16\pi }{5H_0^4}\frac{Q_{\rm rms-PS}^2}{T_0^2}
\biggl[\frac{1}{6\pi }+2n\bar{D}_2^{(2)}(k_{\rm b})\biggr]^{-1}
\biggl[1+2n\bar{h}(k)\biggr]k_{\rm phys}~.
\end{equation}    
The sCDM transfer function is given approximatively by the 
numerical fit~\cite{sugiyama}
\begin{equation}
\label{transfert}
T(k) = \frac{\ln (1+2.34 q)}{2.34 q}\biggl[1+3.89q
+(16.1q)^2+(5.46q)^3+(6.71q)^4\biggr]^{-1/4}, 
\end{equation}
where $q$ and the shape parameter $\Gamma$ can be written as
\begin{equation}
\label{shapepara}
q\equiv k/[(h\Gamma )\mbox{Mpc}^{-1}]
\quad
\quad
\Gamma \equiv \Omega _0 h e^{-\Omega _{\rm b}-
\sqrt{2h}\Omega _{\rm b}/\Omega _0 }~,
\end{equation}
where $\Omega _0$ is the total energy density to critical energy
density ratio and $\Omega _{\rm b}$ represents the baryon
contribution. Or, more explicitly, we take 
$\Omega _0=\Omega _{\rm \Lambda } +\Omega _{\rm m}=\Omega _{\rm
\Lambda }+\Omega _{\rm cdm }+\Omega _{\rm b}$.  We have now normalized
the matter power spectrum to COBE. It is important to realize that the
above procedure only works for the sCDM model since we have used the
Sachs-Wolfe equation. The sCDM matter power spectrum is
depicted in Fig.~\ref{fig1}. The measured power spectrum of the
\emph{IRAS} Point Source Catalogue Redshift Survey (PSCz)~\cite{HT}
has also been displayed for comparison.           
One notices that the effect of the step in $\bar{h}(k)$ is to reduce
the power at small scales which precisely improves the agreement
between the theoretical curves for $n\neq 0$ and the data. Let us
remind at this point that the shape of the function $\bar{h}(k)$ has
not been designed for this purpose and comes from different
(theoretical) reasons. Therefore, it is quite interesting to see that
the power spectrum obtained from our ansatz fits reasonably well the
data. This plot also confirms the result of Ref.~\cite{MRS}, namely
that the number of quanta  must be such that (for sCDM) $1\le n<10$,
i.e., it cannot be too large.

\section{Four-point correlation function for non-vacuum initial states}       

It is time to proceed with the calculation of the four-point
correlation function. Briefly, one has to first perform the
calculation for the state $|\Psi _1 \rangle $, and then generalize it
for $|\Psi _2 \rangle $. 
After some lengthy but straightforward algebra, one finds~\cite{GMS}
\begin{eqnarray}
\label{fourpsi22}
& & \langle \Psi _2(n,k_{\rm b})\vert a_{\ell _1m_1}a_{\ell _2m_2}
a_{\ell _3m_3}a_{\ell _4m_4}\vert \Psi _2(n,k_{\rm b})\rangle
=\frac{\ell_{\rm Pl}^4}{\ell_0^4}\biggl \{ \nonumber \\
& & \ \
(-1)^{m_1+m_2}\biggl[C_{\ell _1}C_{\ell _2}
+2nC_{\ell _1}D_{\ell _2}^{(2)}
+2nC_{\ell _2}D_{\ell _1}^{(2)}
+4n^2F_{\ell _1 \ell _2}^{(2)}\biggr]
\delta _{\ell _1\ell _3}\delta _{\ell _2\ell _4}
\delta _{m_1,-m_3}\delta _{m_2,-m_4}
\nonumber \\
& &
+(-1)^{m_1+m_2}\biggl[C_{\ell _1}C_{\ell _2}
+2nC_{\ell _1}D_{\ell _2}^{(2)}
+2nC_{\ell _2}D_{\ell _1}^{(2)}
+4n^2F_{\ell _1 \ell _2}^{(2)}\biggr]
\delta _{\ell _1\ell _4}\delta _{\ell _2\ell _3}
\delta _{m_1,-m_4}\delta _{m_2,-m_3}
\nonumber \\
& &+(-1)^{m_1+m_3}\biggl[C_{\ell _1}C_{\ell _3}
+2nC_{\ell _1}D_{\ell _3}^{(2)}
+2nC_{\ell _3}D_{\ell _1}^{(2)}
+4n^2F_{\ell _1 \ell _3}^{(2)}\biggr]
\delta _{\ell _1\ell _2}\delta _{\ell _3\ell _4}
\delta _{m_1,-m_2}\delta _{m_3,-m_4}
\nonumber \\
& &
-2n(n+1)E_{\ell _1\ell _2\ell _3\ell _4}^{(2)}
{\cal H}_{\ell _1\ell _2\ell _3\ell _4}^{m_1m_2m_3m_4}
e^{i\pi(\ell _1+\ell _2+\ell _3+\ell _4)/2}
\biggl[1+(-1)^{\ell _1+\ell _3}
+(-1)^{\ell _2+\ell _3}\biggr]\biggr \}~,
\end{eqnarray}      
\vspace*{-15pt}
\begin{eqnarray}
\label{def4b}
F_{\ell _1 \ell _2}^{(2)}
&\equiv &\int _0^{+\infty }{\rm d}\sigma \bar{h}(\sigma )
\frac{{\rm d}}{{\rm d}\sigma }\biggl[D_{\ell _1}^{(1)}
D_{\ell _2}^{(1)}\biggr]
\\
\label{def4bb}
E_{\ell _1\ell _2\ell _3\ell _4}^{(2)}
&\equiv & \int _0^{+\infty }
j_{\ell _1}[k(\eta _0-\eta_{\rm lss})] 
\ldots 
j_{\ell _4}[k(\eta _0-\eta _{\rm lss})]
\bar{h}(k)k^3|f_k|^4\frac{{\rm d}k}{k} 
\end{eqnarray}
We are now in a position to calculate the CMB excess kurtosis. 
In order to establish an analytical formula for it, one just
needs to use the equation linking $a_{\ell m}$ and $\delta T/T$ and 
play with the properties of the spherical harmonics. Explicitly, the
excess kurtosis is defined as           
${\cal K}\equiv \mu_4 -3\mu_2^2~,$
where the second moment has already been introduced and where the
fourth moment, $\mu_4$, of the distribution is defined as
$\mu_4 = \langle K\rangle$ with 
$K \equiv [\frac{\delta T}{T}({\bf e})]^4$. 
An important shortcoming of the previous definition is that the value
of ${\cal K}$ depends on the normalization. It is much more convenient
to work with a normalized (dimensionless) quantity. Therefore, we also
define the normalized excess kurtosis as 
\begin{equation}
\label{defQ}
{\cal Q}\equiv {{\cal K} \over \mu_2^2 }
            =  {\mu_4\over \mu_2^2}-3~,
\end{equation}
which is the one more commonly used in the literature.
\begin{figure}\centerline{\psfig{file=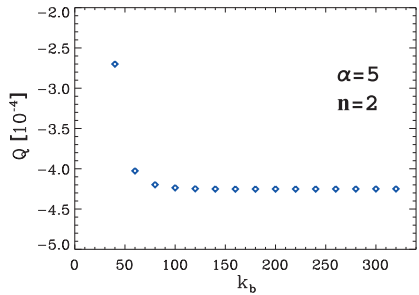,width=6cm}}
\vspace*{-10pt}
\caption{The normalized excess kurtosis parameter ${\cal Q}$ in terms
of the privileged (comoving) wave-number $k_{\rm b}$ for a particular
representative set of parameters: $\alpha= 5$ for the sharpness of the
weight function signaling the privileged scale, and $n= 2$ quanta in
the non-vacuum initial state for the cosmological perturbations.}
\label{fig2}\end{figure}
We have performed the full numerical resolution of the normalized
excess kurtosis parameter which we show in Fig.~\ref{fig2}.  As we see
from it, ${\cal Q}\simeq -4.24 \times 10^{-4}$ is an asymptotic value,
provided we concentrate on the middle and big values of the built-in
scale $k_{\rm b}$. An analytical order of magnitude estimate of this
can be seen in Ref.~\cite{GMS}. 
The fact that the numerical estimate does not depend on
$k_{\rm b}$ is confirmed by the plot, except for small values of the
wave-numbers. In fact, this shows that the quantity ${\cal Q}$ does
not depend very much on the free parameters.  One can see this is true
for parameters $\alpha$ and $n$ since, using the above mentioned
analytical estimate, we find, for $n=1$, ${\cal
Q}\simeq -3.77\times 10^{-4}$ and for $n \to \infty $, ${\cal Q}\simeq
-4.25 \times 10^{-4}$. Since we know that this result does not depend
on the details of the weight function $\bar{h}(k)$, we conclude that
the asymptotic value obtained above is a generic value, at least for
large values of $k_{\rm b}$. In particular, this is true for $k_{\rm
b} \approx 300$ which corresponds to the built-in scale located
roughly at the privileged scale in the matter power spectrum selected
by the redshift surveys of Ref.~\cite{einasto}. Another important
remark is that the excess kurtosis is found to be negative.

\section{Discussion}

The above paragraphs show that the introduction of non-vacuum initial
states leads to a clear prediction for the excess kurtosis
parameter. However, this by itself does not imply that the signature
will be observable. The signal to theoretical noise ratio has to be
considered or, stating it differently, we need to make the comparison
of the signal against the cosmic variance.  The cosmic variance
quantifies the theoretical error coming from the fact that, in
cosmology, observers have only access to one realization of the
$\delta T/T $ stochastic process whereas theoretical predictions are
expressed through ensemble averages. The specific computation of the
cosmic variance for the model under consideration was
performed~\cite{GMS} and found to be some four orders of magnitude
higher than the signal itself. This is no surprise: we are interested
in a non-Gaussian signature of primordial origin, and thus we are
focusing on large angular scales, for which the theoretical
uncertainties are the highest. Shifting to intermediate angular
scales, a stronger signal would be obtained; however, in that case
secondary sources would be more difficult to subtract and thus the
transparency of the effect would be compromised.

In sum, the excess kurtosis is found to be negative and the signal to
noise ratio for the dimensionless excess kurtosis parameter is equal
to $\vert S/N \vert \simeq 4 \times 10^{-4}$, almost independently of
the free parameters of the model. This signature turns out to be
undetectable and therefore we conclude that, subject to current tests, Gaussianity
is a generic property of single field inflationary models.

\section*{Acknowledgements}
I'd like to thank my collaborators J\'er\^ome Martin and Mairi
Sakellariadou for extensive discussions and for hospitality in Paris.
This work was partially financed with funds from {\sc CONICET}, 
{\sc UBA}, {\sc IAP} and {\sc Fundaci\'on Antorchas}.



\begin{thebibliography}{0}

\bibitem{Grishchuk-Martin}L.~P.~Grishchuk and J.~Martin,
Phys.~Rev. {\bf D56}, 1924 (1997),
{\tt gr-qc/9702018}.    

\bibitem{MRS}  J.~Martin, A.~Riazuelo and M.~Sakellariadou,
Phys.~Rev.~D {\bf 61}, 083518 (2000).

\bibitem{silk}
L.~M.~Griffiths, J.~Silk and S.~Zaroubi, Mon.~Not.~R.~Astron.~Soc.
{\bf 324}, 712 (2001).

\bibitem{einasto}
J.~Einasto et al., Nature {\bf 385}, 139 (1997); 
J.~A.~Peacock, Mon.~Not.~R.~Astron.~Soc. {\bf 284}, 885 (1997);
T.~J.~Broadhurst et al., Nature {\bf 343}, 726 (1990).

\bibitem{sugiyama}
N.~Sugiyama, Astrophys.~J.~Suppl. {\bf 100}, 281 (1995).

\bibitem{HT}
A.~J.~S.~Hamilton and M.~Tegmark, {\tt astro-ph/0008392}. 

\bibitem{GMS}
A.~Gangui, J.~Martin and M.~Sakellariadou, 
{\sl Phys.~Rev.~D} {\bf 66}, 083502 (1-23) (2002). [astro-ph/0205202]

\end{thebibliography}
\end{document}